\documentclass[ALICE,manyauthors,nocleardouble]{cernphprep}
\usepackage{graphicx}
\usepackage{dcolumn}
\usepackage{bm}
\usepackage{amstext}
\usepackage{ifpdf}
\ifpdf
  \usepackage[hyperfootnotes=false]{hyperref}
\else
  \usepackage[dvips,hyperfootnotes=false]{hyperref}
\fi
\usepackage[all]{hypcap}
\usepackage{units}
\usepackage{color}
\usepackage{lineno}
\usepackage[numbers,sort&compress]{natbib}
\usepackage[normalem]{ulem} 
\begin{document}%
\newlength{\figwidth}
\setlength{\figwidth}{0.7\columnwidth}
\newcommand\TeV[1]{\unit[#1]{\ensuremath{\mathrm{Te\kern-.1em V}}}}
\newcommand\GeV[1]{\unit[#1]{\ensuremath{\mathrm{GeV}}}}
\newcommand\MeV[1]{\unit[#1]{\ensuremath{\mathrm{MeV}}}}
\newcommand\MeVc[1]{\unit[#1]{\ensuremath{\mathrm{MeV}/c}}}
\newcommand\centiMeter[1]{{\ensuremath\unit[#1]{\mathrm{cm}}}}
\newcommand\centiMeterSquare[1]{{\ensuremath\unit[#1]{\mathrm{cm}^2}}}
\newcommand\microMeterSquare[1]{{\ensuremath\unit[#1]{\mu\mathrm{m}^2}}}
\newcommand\meter[1]{{\ensuremath\unit[#1]{\mathrm{m}}}}
\newcommand\nanoSecond[1]{{\ensuremath\unit[#1]{\mathrm{ns}}}}
\newcommand\SNN[1][]{\ensuremath{s_{\mathrm{NN}}\ifx|#1|\relax\else^{#1}\fi}}
\newcommand\sqrtSNN{\ensuremath{\sqrt{\SNN{}}}}
\newcommand\sqrtSNNTeV[1]{\ensuremath{\sqrtSNN{}=\TeV{#1}}}
\newcommand\sqrtSNNGeV[1]{\ensuremath{\sqrtSNN{}=\GeV{#1}}}
\newcommand\Nch[1][]{\ensuremath{\ifx|#1|\relax N_{\mathrm{ch}}\else%
    N_{\mathrm{ch}#1}\fi}}
\newcommand\barNch[1][]{\ensuremath{\ifx|#1|\relax%
    \mu \else\mu \fi}}
\newcommand\pt{\ensuremath{p_{\mathrm{T}}}}

\newcommand\dNdeta{\ensuremath{\mathrm{d}\Nch{}/\mathrm{d}\eta}}
\newcommand\dNdy{\ensuremath{\mathrm{d}\Nch{}/\mathrm{d}y}}
\newcommand\Npart{\ensuremath{\left\langle
      N_{\mathrm{part}}\right\rangle}}
\newcommand\Namesc[1]{\textsc{#1}}
\newcommand\VZEROsc[1][]{\Namesc{vzero\ifx|#1|\relax\else-#1\fi}}
\newcommand\SPDsc{\Namesc{spd}}
\newcommand\Name[1]{\MakeUppercase{#1}}
\newcommand\ALICE{\Name{alice}}
\newcommand\ATLAS{\Name{atlas}}
\newcommand\CMS{\Name{cms}}
\newcommand\STAR{\Name{star}}
\newcommand\PHOBOS{\Name{phobos}}
\newcommand\BRAHMS{\Name{brahms}}
\newcommand\RHIC{\Name{rhic}}
\newcommand\SPS{\Name{sps}}
\newcommand\AGS{\Name{ags}}
\newcommand\LHC{\Name{lhc}}
\newcommand\CERN{\Name{cern}}
\newcommand\GEANT{\Name{geant3}}
\newcommand\HIJING{\Name{hijing}}
\newcommand\FMD[1][]{\Name{fmd}\ifx|#1|\relax\else#1\fi}
\newcommand\VZERO[1][]{\Name{vzero\ifx|#1|\relax\else-#1\fi}}
\newcommand\TZERO[1][]{\Name{t{\small0}\ifx|#1|\relax\else-#1\fi}}
\newcommand\TPC{\Name{tpc}}
\newcommand\SPD{\Name{spd}}
\newcommand\ZDC{\Name{zdc}}
\newcommand\ZEM{\Name{zem}}
\newcommand\AuAu{Au--Au}
\newcommand\PbPb{Pb--Pb}
\newcommand{\figabbrevname}{Fig.}
\newcommand{\figsabbrevname}{Figs}
\newcommand{\figfullname}{Figure}
\newcommand{\figsfullname}{Figures}
\renewcommand{\figurename}{Fig.}
\renewcommand{\tablename}{Table}
\newcommand\tabref[1]{\tablename~\ref{#1}}
\newcommand\TODO[1]{{\it\color[rgb]{.8,.2,.2} #1}}
\newcommand{\blu}[1]{\textbf{\textcolor{blue}{#1}}}
\renewcommand{\xout}[1]{\textcolor{red}{\sout{#1}}}
\newcommand{\ask}[1]{\textcolor{magenta} {#1} }

%
\begin{titlepage}
\PHnumber{2013-045}                 
\PHdate{March 22, 2013}              
%
%
\title{Centrality dependence of the pseudorapidity density
  distribution for charged particles in \PbPb{} collisions at
  $\mathbf{\sqrtSNNTeV{2.76}}$}
\ShortTitle{Centrality dependence of the pseudorapidity density}   
%
\Collaboration{ALICE Collaboration%
         \thanks{See Appendix~\ref{app:collab} for the list of collaboration
                      members}}
\ShortAuthor{ALICE Collaboration}      
\begin{abstract}
  We present the first wide-range measurement of the charged-particle pseudorapidity
  density distribution, for different centralities (the 0--5\%, 5--10\%, 10--20\%, and
  20--30\% most central events) in \PbPb{} collisions at
  \sqrtSNNTeV{2.76} at the \LHC{}. The measurement is performed using the full coverage
  of the \ALICE{} detectors, $-5.0 < \eta < 5.5$, and employing a special analysis
  technique based on collisions arising from \LHC{} `satellite' bunches. We present the
  pseudorapidity density as a function of the number of participating nucleons as well
  as an extrapolation to the total number of produced charged particles
  ($\Nch{} = 17165 \pm 772$ for the 0--5\% most central collisions). From the
  measured \dNdeta{} distribution we derive the rapidity density distribution,
  \dNdy{}, under simple assumptions. The rapidity density distribution is found to be
  significantly wider than the predictions of the Landau model. We assess the validity
  of longitudinal scaling by comparing to lower energy results from \RHIC{}. Finally
  the mechanisms of the underlying particle production are discussed based on a
  comparison with various theoretical models.

\end{abstract}

\end{titlepage}
\setcounter{page}{2}
\section{Introduction}

There exists much evidence that, under the extreme
conditions of unprecedented temperature and energy density created in
ultra-relativistic heavy-ion collisions, matter is in a deconfined state known as the
quark--gluon plasma \cite{Arsene:2004fa,Back:2004je,Adams:2005dq,Adcox:2004mh}.
A new era  in the study of these collisions began with the production
of \PbPb{} collisions at a center-of-mass energy per nucleon pair
\sqrtSNNTeV{2.76} at the \CERN{} \LHC{}.

The charged-particle pseudorapidity density generated in heavy-ion collisions depends on the
particle production mechanisms as well as on the initial energy density.
Studying the dependence of the pseudorapidity density on collision centrality will
improve our understanding of the role of hard scattering and
soft processes contributing to the production of charged particles
(e.g. parton saturation \cite{Gribov:1984tu}). Moreover, extending the measurement to a wide
pseudorapidity range enables investigating the physics of
the fragmentation region by comparing the extrapolation of this
data to lower energy data from \RHIC{}
\cite{Alver:2010ck,Back:2005hs} to test whether longitudinal
scaling of the pseudorapidity density persists up to \LHC{} energies.
 
In this Letter we present the first \LHC{} measurement over a
wide pseudorapidity range of the centrality dependence of the charged-particle
pseudorapidity density, \dNdeta{}, utilizing the \ALICE{} detector. The employed method
relies on using so-called `satellite' bunch collisions and is based on
measurements from three different \ALICE{} sub-detectors. This method is applicable
for the 30\% most central events where the trigger efficiency for these `satellite' collisions
remains high. These measurements extend considerably the former results obtained at the \LHC{}
\cite{Aamodt:2010pb,Chatrchyan:2011pb,ATLAS:2011ag} and can be compared to the
wealth of results on the charged-particle pseudorapidity density from lower energy \AuAu{}
collisions at \RHIC{} \cite{Alver:2010ck,Bearden:2001xw,Bearden:2001qq} as well
as model calculations.

\section{Experimental setup}

\begin{figure*}[htbp]
  \centering
  \includegraphics[keepaspectratio,width=\textwidth]{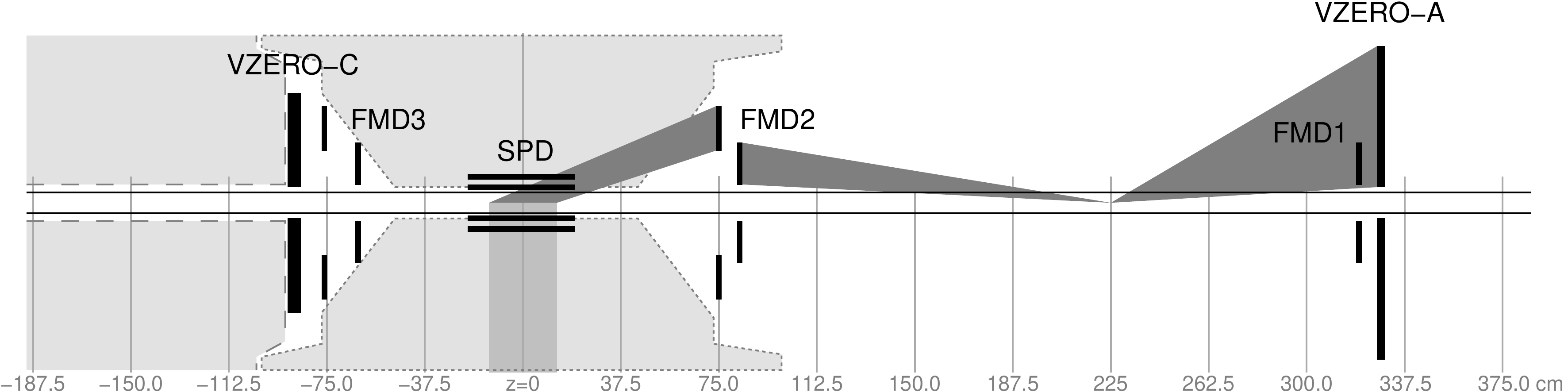}
  \caption{Schematic drawing (not to scale) of the cross-section of the sub-detectors
    used in this analysis and the midpoints of the locations of the nominal and
    `satellite' interaction points. The long-dashed line designates a region of
    dense material designed to absorb all particles except muons. The
    short-dashed line indicates the region of the \ALICE{}
    inner tracking system, which has dense material for its services on the
    surfaces near \FMD[2]{} and \FMD[3]{}. The area between
    \FMD[2]{}, \FMD[1]{} and \VZERO[A]{} contains only the
    beryllium beam pipe. The dark gray shaded areas denote the paths particles
    would follow from $z = \centiMeter{0}$ and $z = \centiMeter{225}$ to \FMD[2]{}
    and \VZERO[A]{} such that it is evident which material they would traverse.}
  \label{fig:layout}
\end{figure*}

A detailed description of the \ALICE{} detector can be found in
\cite{Aamodt:2008zz}. In the following, we will briefly describe the
detectors used in this analysis, namely the Silicon Pixel Detector (\SPD{}), the
Forward Multiplicity Detector (\FMD{}), the \VZERO{}, and the Zero Degree Calorimeter
(\ZDC{}) (see \figabbrevname~\ref{fig:layout}). 

The \SPD{} is the innermost element of the \ALICE{} inner tracking system \cite{Aamodt:2008zz}. It
consists of two cylindrical layers of hybrid silicon pixel assemblies
positioned at radial distances of $3.9$ and $\centiMeter{7.6}$ from
the beam line, with a total of $9.8\times10^6$ pixels of size
$\microMeterSquare{50\times425}$, read out by 1200 electronic
chips. The \SPD{} coverage for particles originating from the nominal interaction point at the center
of the detector is $|\eta|<2$ and $|\eta|<1.4$ for the inner and outer
layers, respectively.
  
The \VZERO{} detector \cite{Abbas:2013taa} consists of two arrays of 32 scintillator tiles
(4 rings of increasing radii each with 8 azimuthal sectors)
placed at distances of $\meter{3.3}$ (\VZERO[A]{}) and $\meter{-0.9}$ (\VZERO[C]{}) from the
nominal interaction point along the beam axis, covering the full azimuth within $2.8 <
\eta < 5.1$ and $-3.7 < \eta < -1.7$, respectively. Both
the amplitude and the time of the signal in each scintillator are recorded.

The \ZDC{} measures the energy of spectator (non-interacting) nucleons
in two identical sets of detectors, located at $\pm\meter{114}$ from the
interaction point along the beam axis \cite{Aamodt:2008zz}. Each set consists
of two quartz fiber sampling calorimeters: a neutron calorimeter positioned
between the two \LHC{} beam pipes down-stream of the first \LHC{} dipole which
separates the two charged-particle beams and a proton calorimeter positioned
externally to the beam pipe containing bunches moving away from the interaction point.
The \ZDC{} energy resolution at the Pb beam energy
is estimated to be 20\% and 25\% for the neutron and proton
calorimeters, respectively. The \ZDC{} system is completed by two
Zero-degree Electro-Magnetic (\ZEM{}) calorimeters placed at
$+\meter{7.5}$ from the interaction point along the beam direction
\cite{Aamodt:2008zz}. They cover the pseudorapidity range
between $4.8$ and $5.7$ and thus measure the energy of
particles emitted at very small angles with respect to the beam axis.

The \FMD{} \cite{Christensen:2007yc} is composed of three sub-detectors: \FMD[1]{}, \FMD[2]{}, and \FMD[3]{}. \FMD[2]{} and \FMD[3]{} consist of
an inner and an outer ring of silicon strip sensors, while \FMD[1]{} consists of only an inner ring. The
inner and outer rings have internal radii of $\centiMeter{4.2}$ and
$\centiMeter{15.4}$ and external radii of $\centiMeter{17.2}$ and
$\centiMeter{28.4}$, respectively, with full azimuthal coverage.
Each ring is sub-divided into 512 or 256
radial strips and 20 or 40 azimuthal sectors for inner and outer
rings, respectively.  For collisions at the nominal interaction point
the pseudorapidity coverage is $-3.4<\eta<-1.7$ (\FMD[3]{}) and
$1.7<\eta<5.0$ (\FMD[2]{} and \FMD[1]{}). Each sub-ring has 10240 channels
resulting in a total of 51200 channels.

\section{Data sample and analysis approach}

The analysis presented in this Letter is based on \PbPb{} collision data
at \sqrtSNNTeV{2.76} taken by \ALICE{} in 2010.

Results in the region of $|\eta|<2$ are obtained from
a tracklet analysis using the two layers of the \SPD{}.
The analysis method and the used data sample
are identical to the ones described in
\cite{Aamodt:2010cz}, but extending
the pseudorapidity range of the \SPD{} detector
by using collisions occurring within $\pm \centiMeter{13}$
(rather than $\pm \centiMeter{10}$) from the nominal
interaction point along the beam axis.

The measurement of multiplicity in the region $|\eta|>2$ is carried out
using the \FMD{} and \VZERO{}. The main challenge in the analysis of
these data is the correction for secondary particles
produced by primary particles interacting with the detector material. While the low
material density in the \ALICE{} central barrel
limits the number of secondary particles,
for $|\eta|>2$ dense material -- such
as services, cooling tubes, and read-out cables -- is present. This material
causes a very large production of secondary particles, in
some cases up to twice the number of primary particles
as obtained from Monte Carlo studies. Furthermore,
the geometry and segmentation of the two detectors do not allow
for the rejection of secondary particles through tracklet reconstruction
and therefore the analysis depends strongly on Monte Carlo driven corrections.
In order to reduce systematic effects arising from these large correction
factors, a special analysis technique was developed.
It relies on the so-called `debunching' effect which occurs during the
injection and acceleration of the beams inside the \LHC{} ring \cite{Welsch:2012zzf}. 
Due to the way the beams are injected and transferred to the \LHC{},
a small fraction of the beam can be captured in unwanted
RF buckets which creates so-called `satellite' bunches spaced by $\nanoSecond{2.5}$.
Thus crossings of the `satellite' bunches of one beam with the main bunches of the
opposite beam produce `satellite' interactions with vertices spaced by $\centiMeter{37.5}$
in the longitudinal direction (see \figabbrevname~\ref{fig:layout}).
These interactions provide the opportunity to avoid the
large amount of material traversed by particles coming from
the nominal vertex and to extend the pseudorapidity range of the \FMD{} and \VZERO{}.
Interactions with vertices from $\centiMeter{-187.5}$ to $\centiMeter{375}$ are
used in this analysis. Furthermore, \FMD[3]{} and \VZERO[C]{} are surrounded by dense material
and, therefore, only the \FMD[1]{}, \FMD[2]{}, and \VZERO[A]{} were used. For
`satellite' collisions in the range of $75, 102.5, \ldots,
\centiMeter{300}$ from the nominal interaction point along the beam axis,
the only material between the \VZERO[A]{}, \FMD1{}, \FMD2{} and the
interaction point is the beryllium beam pipe, resulting in
a reduction of the number of secondary particles by more than a factor of two in Monte Carlo simulations
and consequently much smaller corrections. For vertices with $z > \centiMeter{300}$
and $z < \centiMeter{75}$ other detector material has an increasing influence on the
measurement such that for vertices with $z < \centiMeter{37.5}$ only \FMD1{} and
the inner ring of \VZERO[A]{} are used. An additional advantage of this analysis method is
the possibility for a data-driven calibration of the detector response using
`satellite' collisions for which the pseudorapidity coverage of the \VZERO{}
overlaps with the nominal \SPD{} acceptance, as it will be explained in the following.

Due to the fact that the `satellite' collision vertices fall outside
the normal range around the nominal interaction point, the standard
\ALICE{} trigger and event selection \cite{Aamodt:2010pb} is inapplicable.
Therefore a special trigger imposing a lower cut of $100$ fired chips on both layers
of the \SPD{} was used. The trigger was verified to be fully efficient for the centrality
range covered by the present analysis. This was done by inspecting the distribution of
the number of fired \SPD{} chips as a function of the event centrality.
The triggered events are then further selected based on the \ZDC{} timing information, so that 

$$
 \frac{\left(\Delta T - n\times\nanoSecond{2.5}\right)^2}{
   (\sigma_{\mathrm{\Delta T}})^2} +
 \frac{\left(\Sigma T - n\times\nanoSecond{2.5}\right)^2}{
   (\sigma_{\mathrm{\Sigma T}})^2} < 1\quad,
$$

\noindent where $\Delta T$ and $\Sigma T$ are the difference and sum of the arrival times
(relative to the crossing time of the main bunches) of the signals in the two \ZDC{}
calorimeters, respectively, and $\sigma_{\mathrm{\Delta T}}=\nanoSecond{1.32}$
and $\sigma_{\mathrm{\Sigma T}}=\nanoSecond{2.45}$ are the corresponding resolutions.
$n$ is the index of the `satellite' interaction point, such that $n = 0$ denotes an interaction
at the nominal interaction point. More details
on the event selection can be found in \cite{Abelev:2013qoq}.
It is worth noting that the crossing angle between the beams was zero during the \PbPb{} data
taking in 2010 which naturally enriched the data sample with `satellite' collisions.
The rate of the `satellite' collisions
is about three orders of magnitude lower than the rate of the nominal collisions and therefore,
in order to accumulate a sufficient amount of events, the analysis was performed with all
`satellite' collisions from the entire 2010 data sample. The acquired
statistics is distributed unevenly among the different `satellite' vertices and varies
from one thousand to twelve thousand events per vertex.

Similarly to the trigger and event selection, the standard centrality selection
based on \VZERO{} \cite{Abelev:2013qoq} can not be used in the analysis of
the `satellite' collisions.
Given the fact that the \ZDC{} and \ZEM{} are positioned very far away from the
nominal interaction point, they are best suited for the characterization of
`satellite' collisions. The event sample is split into four centrality classes
(0--5\%, 5--10\%, 10--20\%, and 20--30\%) based on the energy deposited by
spectator nucleons in the \ZDC{} and by particles emitted at small angles with respect to the beam axis in the
\ZEM{}. The number of spectator nucleons and,
therefore, their deposited energy decreases for more central
events while the inverse is true for particles emitted at small angles with respect to the beam axis. One can therefore
define centrality cuts based on this anti-correlation. In order to
match this estimator to the standard \ALICE{} centrality selection, the correlation between the
\ZDC{} versus \ZEM{} and \VZERO{} signal for events near the nominal interaction
point is determined \cite{Abelev:2013qoq}. This method is only reliable
in the centrality range 0--30\% where the trigger is also fully efficient; this defines the centrality range
for the presented measurement. To reduce the residual bias arising from the position of
the interaction point, the \ZEM{} signal is scaled as a function of `satellite' vertex.
The scaling factors are obtained by a linear fit to the \ZDC{} versus \ZEM{} anti-correlation.
They are found to be between $0.96$ and $1.04$ for vertices from $\centiMeter{-187.5}$ to
$\centiMeter{225}$ and about $0.86$ for the farthest vertex at $\centiMeter{375}$.

The \FMD{} and \VZERO{}
are used to extract the multiplicity independently in the same $\eta$ acceptance. The \FMD{} records the
energy loss of particles that traverse each silicon strip.
The first step in the analysis is to apply a minimum cut on the measured energy to neglect
strips considered to have only electronics noise. Due to the
incident angle of the particles impinging on the detector, the
energy loss signal of one particle may be spread out over more than one strip. The next step
in the analysis is therefore to cluster individual strip signals
corresponding to the energy of a single particle.
This is accomplished by adding the strip signals which are below a clustering threshold
to neighboring strips which have a larger signal if one exists. The finite resolution of the \FMD{} also allows
for more than one particle to traverse a single strip.
The number of charged particles per strip is then determined using a statistical
approach where the mean number of particles per strip, $\barNch[,\FMD]$, over a
region of 256 strips (64 strips radially $\times$ 4 strips azimuthally) is
estimated assuming a Poisson distribution, such that
$\barNch[,\FMD] = -\ln\left( N_{\mathrm{E}} / N_{\mathrm{S}} \right)$,
where $N_{\mathrm{E}}$ is the number of strips with no hits and $N_{\mathrm{S}}$ is the total
number of strips (256 here) in the defined region.  To get the average number of
particles per hit strip, a correction of $c = \tfrac{\barNch[,\FMD]}{1 -
  e^{-\barNch[,\FMD]}}$ is applied to each hit strip in the region.
Next, the data are corrected for the acceptance at a given interaction
point, and the inclusive charged-particle count is converted to the
number of primary charged particles by means of an interaction-point
and centrality-dependent response map. These response maps are based
on \GEANT{} \cite{Brun:1994aa} Monte Carlo simulations using the \HIJING{} event generator
\cite{Wang:1991hta} and relate the number of generated primary
charged particles in a given $(\eta,\varphi)$ bin (bins are of size $0.05$
in $\eta$ and $\pi/10$ in $\varphi$) to the total number of
charged particles reduced by the detector efficiency in the same bin. The response maps are
highly sensitive to the accuracy of the experimental description in
the simulation, and are therefore the largest source of the
systematic error on the results from the \FMD{}.

In order to calculate the charged-particle density in the \VZERO{} detector,
the Monte Carlo simulation described above is used to relate the observed
signal to the number of primary charged particles within the acceptance of a
given \VZERO{} ring. The relation is given by $A(z,i) = \alpha(z,i) \Nch{}(\eta(z,i))$, where
$i$ is the ring index and $z$ is the longitudinal position of the interaction point.
$A$ is the \VZERO{} signal amplitude, \Nch{} is the number of primary charged particles
in the \VZERO{} ring's acceptance from the given interaction point, and $\alpha$ is the conversion
factor between $A$ and \Nch{} determined from the Monte Carlo simulation. In order to
minimize the dependence on
the simulation and perform a data-driven analysis, the \VZERO{} response is
calibrated using reference `satellite'
vertices, $z_{\mathrm{r}}$, between $\centiMeter{225}$ and $\centiMeter{375}$ for which the pseudorapidity
coverage of the \VZERO{} rings lies inside $|\eta|<2$, i.e. overlapping the range
of the \SPD{} at the nominal interaction point.
In this way the charged-particle pseudorapidity density in a given ring of
the \VZERO{} detector and for a given interaction point, in the range of
$\centiMeter{-187.5} \leq z \leq \centiMeter{375}$, is obtained as:

$$
\frac{\mathrm{d}N_{\mathrm{ch}}^{\VZEROsc{}}}{\mathrm{d}\eta}(\eta(z,i)) = \frac{\mathrm{d}N_{\mathrm{ch}}^{\SPDsc{}}}{\mathrm{d}\eta}(\eta(z_{\mathrm{r}},i))\frac{\alpha(z_{\mathrm{r}},i)}{\alpha(z,i)}\frac{A(z,i)}{A(z_{\mathrm{r}},i)}\;,
$$ 

\noindent where $\mathrm{d}N_{\mathrm{ch}}^{\SPDsc{}}/\mathrm{d}\eta$ is the charged-particle pseudorapidity density measured by the \SPD{},
$z_{\mathrm{r}}$ is the longitudinal position of the reference vertex and $\eta$ is
the pseudorapidity corresponding to the chosen vertex and \VZERO{} ring.
The factors $\alpha$ represent the full detector response including secondary particles, light
yield per particle, and electronics response of the \VZERO{}, and
are checked to be constant for the selected range of `satellite' vertices.

Finally, a small correction (up to 1\%) is applied to the \VZERO{} data points arising from
a residual bias in the method determined from Monte Carlo simulations by comparing the final
reconstructed \dNdeta{} distribution after combining the results from all vertices to the
Monte Carlo input \dNdeta{} distribution.

\section{Systematic errors} 

\begin{table}[htbp]
  \centering
  \caption{List of the considered systematic errors.}
  \begin{tabular}{ccr}
    \hline
    \hline
    Detector  & Source                                     & Error    \\
    \hline 
    \hline 
    Common    & Centrality                                 & 1--2\%   \\
    \hline
              & Background subtraction                     & 0.1--2\% \\
    \SPD{}    & Particle composition                       & 1\%      \\
              & Weak decays                                & 1\%      \\
              & Extrapolation to zero momentum             & 2\%      \\
    \hline
    \FMD{} \& & Material budget                            & 4\%      \\
    \VZERO{}  & \ZEM{} scaling                             & 1--4\%   \\
    \hline
              & Particle composition, spectra, weak decays & 2\%      \\
    \FMD{}    & Variation of cuts                          & 3\%      \\
              & Analysis method                            & 2\%      \\
    \hline
    \VZERO{}  & Variation between rings                    & 3\%      \\
              & Calibration by \SPD{}                      & 3--4\%   \\
    \hline
    \hline
  \end{tabular}
  \label{tab:systematic_errors}
\end{table}

\tabref{tab:systematic_errors} summarizes the various contributions to
the systematic errors for each of the three detectors used, as well as the
common contribution arising from the uncertainty in the centrality determination.
The latter is assessed by comparing the \SPD{} results obtained with the standard
approach based on the total \VZERO{} amplitude and the \ZDC{} versus \ZEM{}
anti-correlation. The details of \tabref{tab:systematic_errors} are explained
in the following paragraphs.

A related source of systematic error
which affects `satellite' vertices and hence only the \FMD{} and \VZERO{} analyses is
the uncertainty of the \ZEM{} scaling factors. This was evaluated by varying the \ZEM{}
scaling factors between the values obtained via a linear fit to the \ZDC{} versus \ZEM{} anti-correlation
and the values which give the appropriate number of events
in each centrality bin (i.e. the 0--5\% bin should have the same number of events as the 5--10\% bin and
half the number of events of the 10--20\% and 20--30\% bins) and studying the effect
on the final values. The influence of the particle composition, the particle spectra and the relative fraction
of weak decays of $\Lambda$ and $K^0_s$ are studied by modifying these quantities within the Monte Carlo
simulation in order to match the measured particle spectra and yields \cite{Abelev:2012wca,ALICE:2013xaa}.
The uncertainty due to the description of the material budget
in the region concerned by the analysis was estimated by varying the contribution
of secondary particles from interactions in the detector material by 10\%.

For the \FMD{}, two detector-specific contributions to the systematic error are considered.
First, the noise cut and clustering threshold, determining which strips have no or partial
signals from particles, are varied by $\pm 10\%$. This was found from simulations to be the
range in which the probability to identify two particles as one and a single particle as
multiple particles is minimal. The effect of these variations on
the final result is a component of the systematic error. Secondly, an alternative method is used to determine
the \FMD{} multiplicity. The method using Poisson statistics is compared to a method using the distributions
of deposited energy in the \FMD{}. The difference between the results obtained by the
two methods (2\%) is an additional component of the systematic error. 

The systematic error in the \VZERO{} measurement stems mainly from the uncertainty of the \SPD{} results
used to calibrate the \VZERO{} response. The systematic error related to the \SPD{} analysis is
described in detail in \cite{Aamodt:2010cz}
and is the basis of the uncertainty on the \VZERO{} calibration.
A further contribution to the systematic error is assessed by taking into account
the variation between the results from various \VZERO{} rings at different `satellite' vertices that
cover the same or close pseudorapidity ranges.

\section{Results}

\begin{figure}[htbp]
  \centering
  \includegraphics[keepaspectratio,width=\figwidth]{%
    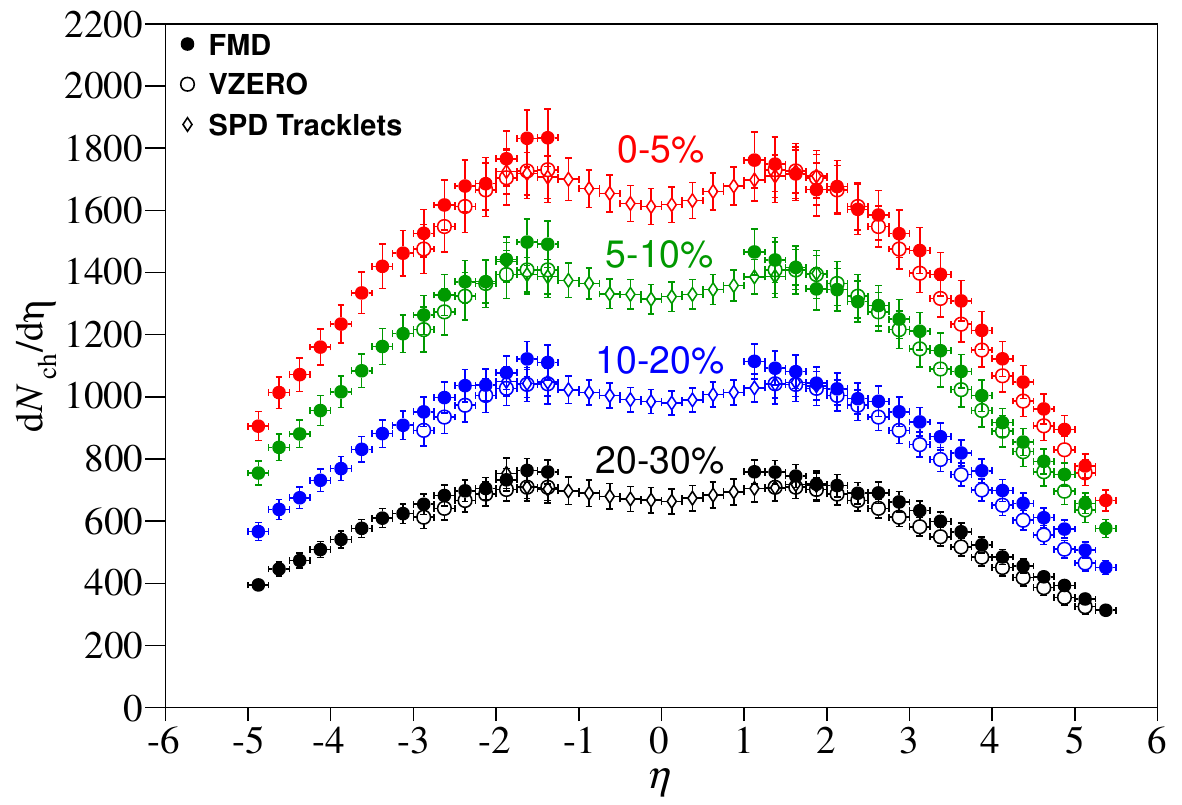}
  \caption{\dNdeta{} per centrality bin from each of the three detectors used. The error bars
  correspond to the total statistical and systematic error.}
  \label{fig:individual_dndeta}
\end{figure}

\begin{figure}[htbp]
  \centering
  \includegraphics[keepaspectratio,width=\figwidth]{%
    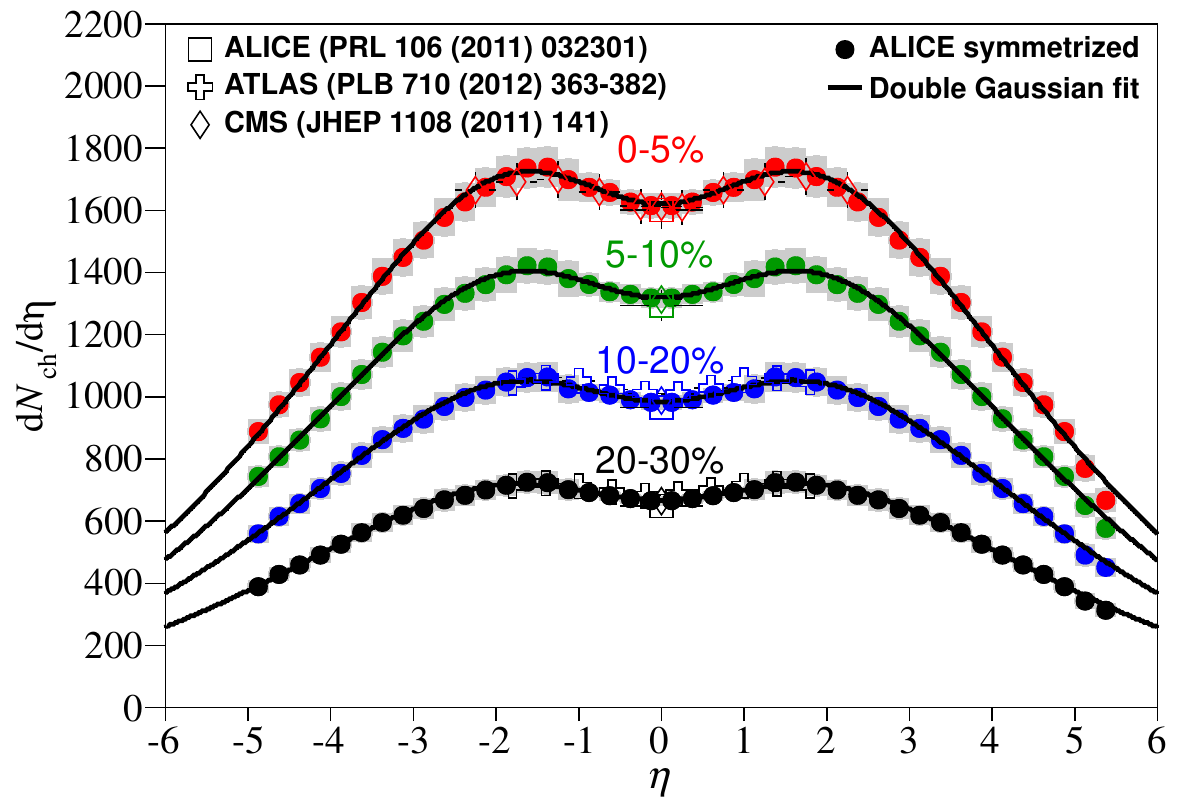}
  \caption{Combined \dNdeta{} result per centrality bin. The error bars (gray boxes) show
  the total statistical and systematic error of the combined result. The open squares
  indicate the previously published \ALICE{} result near mid-rapidity \protect\cite{Aamodt:2010cz}.
  Published results from other \LHC{} experiments \protect\cite{Chatrchyan:2011pb,ATLAS:2011ag} which
  have the same centrality as the \ALICE{} measurement are also shown.}
  \label{fig:combined_dndeta}
\end{figure}

\begin{figure}[htbp]
  \centering
  \includegraphics[keepaspectratio,width=\figwidth]{%
    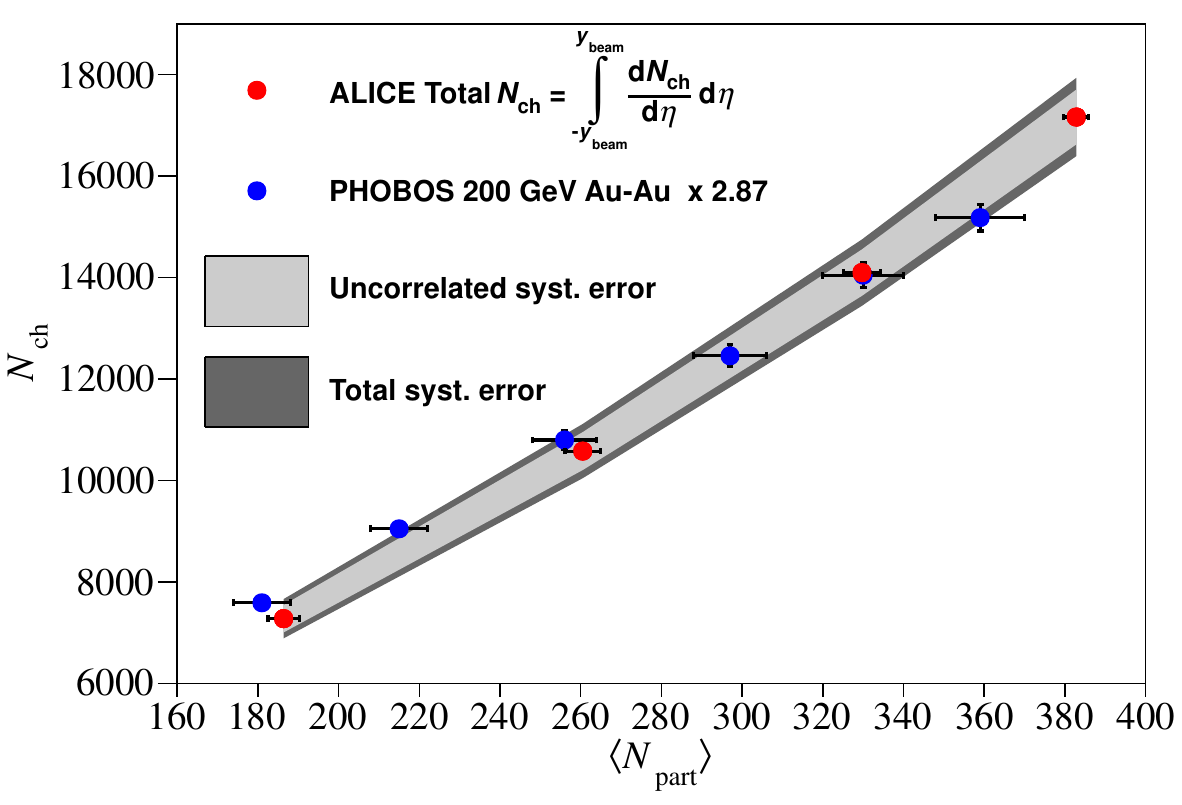}
  \caption{Extrapolation to the total number of produced charged particles as a function of the number
    of participating nucleons. The light-gray band represents the uncorrelated errors from the extrapolation fit
    while the dark-gray band shows the increase to the total systematic errors which includes the common error
    coming from the uncertainty in the centrality
    estimation. The lower energy data from PHOBOS \protect\cite{Alver:2010ck} was scaled by the average
    number of charged particles per participant with $\Npart{} > 180$ found in the \ALICE{} measurements divided by the same
    quantity found in the PHOBOS measurements.}
  \label{fig:nch_integral}
\end{figure}

\figsfullname~\ref{fig:individual_dndeta} and \ref{fig:combined_dndeta} show the
resultant charged-particle pseudorapidity density from each of the three detectors
individually and combined, respectively. The combined distribution is
computed as the average value of \dNdeta{} between the various detectors weighted by
the systematic errors that are not common to the detectors (the statistical errors are
negligible in comparison to the systematic errors). The error obtained from
this weighting is then summed in quadrature with the common systematic errors. Finally,
the distribution is symmetrized around $\eta = 0$ in the range of $|\eta| < 5$ by
computing the average of \dNdeta{} at positive and negative $\eta$
values weighted by their systematic errors. This positive--negative asymmetry
varies between 1\% and 8\%. The resultant distribution
is in agreement with those measured by \ATLAS{} \cite{ATLAS:2011ag} and \CMS{} \cite{Chatrchyan:2011pb}.
The lines on \figabbrevname~\ref{fig:combined_dndeta} represent fits to the following function: 

$$
A_{\mathrm{1}} e^{-\frac{\eta^2}{2\sigma_{\mathrm{1}}^2}} - A_{\mathrm{2}} e^{-\frac{\eta^2}{2\sigma_{\mathrm{2}}^2}}\quad,
$$

\noindent that is the difference of two Gaussians centered at $\eta=0$ and having
amplitudes $A_{\mathrm{1}}$, $A_{\mathrm{2}}$ and widths $\sigma_{\mathrm{1}}$, $\sigma_{\mathrm{2}}$. For the 0--5\% bin, $A_{\mathrm{1}} = 2102 \pm 105$,
$A_{\mathrm{2}} = 485 \pm 99$, $\sigma_{\mathrm{1}} = 3.7 \pm 0.1$, and $\sigma_{\mathrm{2}} = 1.2 \pm 0.2$. The values of $A_{\mathrm{1}} / A_{\mathrm{2}}$,
$\sigma_{\mathrm{1}}$, and $\sigma_{\mathrm{2}}$ are the same for each measured centrality bin within errors. This function
describes the data well within the measured region and gives the best fit among multiple functions used to extrapolate
the distribution to $\pm y_{\mathrm{beam}}$ ($y_{\mathrm{beam}} = 7.99$ at \sqrtSNNTeV{2.76}) in order to obtain
the total charged-particle yield. The results of the extrapolation are summarized in
\tabref{tab:summary} and \figabbrevname~\ref{fig:nch_integral}. The extrapolation is
performed using four different fit functions: the Gaussian function mentioned earlier,
a trapezoidal function from \cite{Alver:2010ck}, a function composed of a hyperbolic cosine
and exponential also from \cite{Alver:2010ck}, as well as a Bjorken inspired function composed
of a central plateau with Gaussian tails. The central value of the extrapolation is derived
from the trapezoidal function as little yield is expected beyond $y_{\mathrm{beam}}$. The quoted
errors include the variation of the fit parameters due to the measurement uncertainties as well
as the deviations between the four fit functions
in the region beyond the pseudorapidity range covered by the experimental data.
The total number of produced charged particles as a function of the number of participating
nucleons shows a similar behavior as at lower energies when scaled to have the same
average number of charged particles per participant (see \figabbrevname~\ref{fig:nch_integral}).

\begin{table}[htbp]
  \centering
  \caption{The number of participants \Npart{} estimated from the Glauber model
    \protect\cite{Abelev:2013qoq} and the total charged-particle yield in the measured region
    ($-5.0 < \eta < 5.5$) and extrapolated to $\pm y_{\mathrm{beam}}$ for different centrality fractions.}
  \begin{tabular}{r@{$-$}lr@{$\pm$}lr@{$\pm$}lr@{$\pm$}lr@{$\pm$}l}
    \hline
    \hline
    \multicolumn{2}{l}{Centrality [\%]} & 
    \multicolumn{2}{c}{\Npart{}} & 
    \multicolumn{2}{c}{$\Nch[,-5.0 < \eta < 5.5]$} & 
    \multicolumn{2}{c}{$\Nch[,|\eta| \leq y_{\mathrm{beam}}]$} \\ 
    \hline
    0 & 5 & 382.8&3.1 & \phantom{100}14963&666 & \phantom{100}17165&772\\ 
    5 & 10  & 329.7&4.6 & 12272&561 & 14099&655\\
    10 & 20 & 260.5&4.4 & 9205&457 & 10581&535\\
    20 & 30 & 186.4&3.9 & 6324&330 & 7278&387\\
    \hline
    \hline
  \end{tabular}
  \label{tab:summary}
\end{table}

\begin{figure}[htbp]
  \centering
  \includegraphics[keepaspectratio,width=\figwidth]{%
    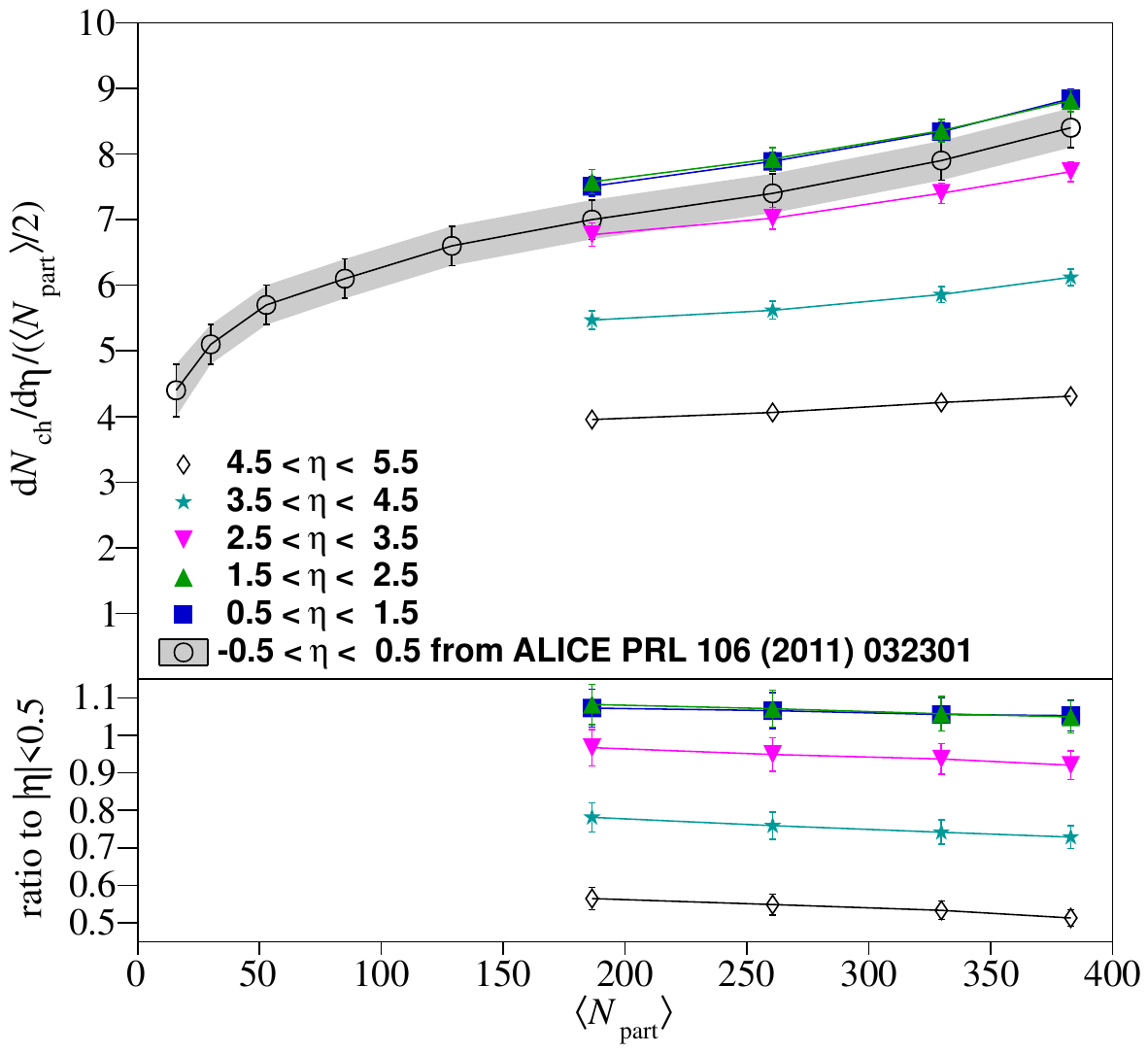}
  \caption{$\dNdeta{}/($\Npart{}$/2)$ as a function of \Npart{} for different $\eta$
    ranges. The lower panel shows the ratio of each distribution to the published
    distribution from data with $|\eta| < 0.5$.}
  \label{fig:dndeta_npart}
\end{figure}

In \figabbrevname~\ref{fig:dndeta_npart} we present the charged-particle pseudorapidity density
per participating nucleon pair, \Npart{}$/2$, as a function of \Npart{} for different
pseudorapidity ranges. The lower panel of \figabbrevname~\ref{fig:dndeta_npart}
shows no strong evolution in the shape of the pseudorapidity density distribution as a function of
event centrality for the 30\% most central events.

\begin{figure}[htbp]
  \centering
  \includegraphics[keepaspectratio,width=\figwidth]{%
    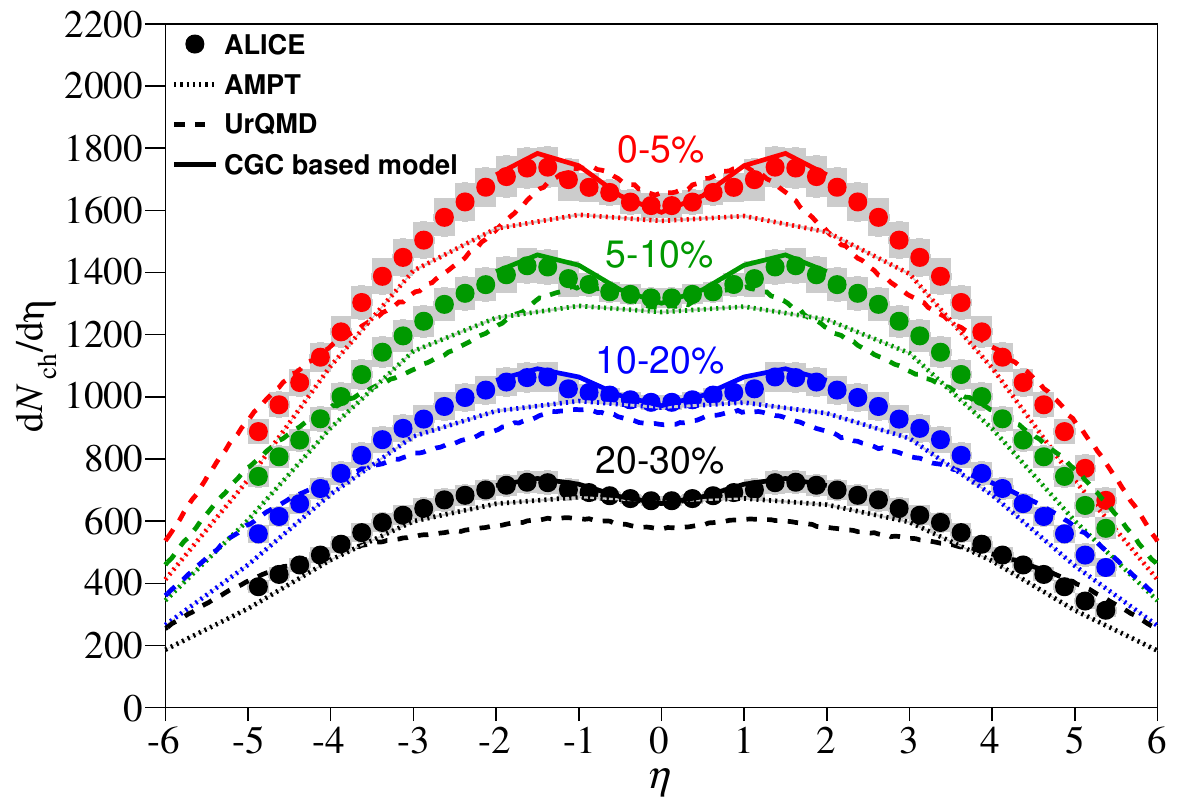}  
  \caption{\dNdeta{} per centrality class compared to model predictions
    \protect\cite{ALbacete:2010ad,Albacete:2011fw,Mitrovski:2008hb,Lin:2004en,Xu:2011fi}.}
  \label{fig:dndeta_model}
\end{figure}

We have compared our measurement to three theoretical models which predict the
pseudorapidity density -- a Color Glass Condensate (CGC) based model
\cite{ALbacete:2010ad,Albacete:2011fw}, the UrQMD model \cite{Mitrovski:2008hb}, and
the AMPT model \cite{Lin:2004en} as tuned in \cite{Xu:2011fi}.
As seen in \figabbrevname~\ref{fig:dndeta_model}, in its limited pseudorapidity range
($|\eta| < 2$) the CGC based model has a similar shape to the measured result.
The UrQMD model gives a reasonable description of the region $|\eta|>4$ and the shape
at mid-rapidity, but is unable to describe the overall level of the pseudorapidity density
as well as most of the shape. The AMPT model does reproduce the level at mid-rapidity
as it was tuned for, but fails to reproduce the overall shape.

\begin{figure}[htbp]
  \centering
  \includegraphics[keepaspectratio,width=\figwidth]{%
    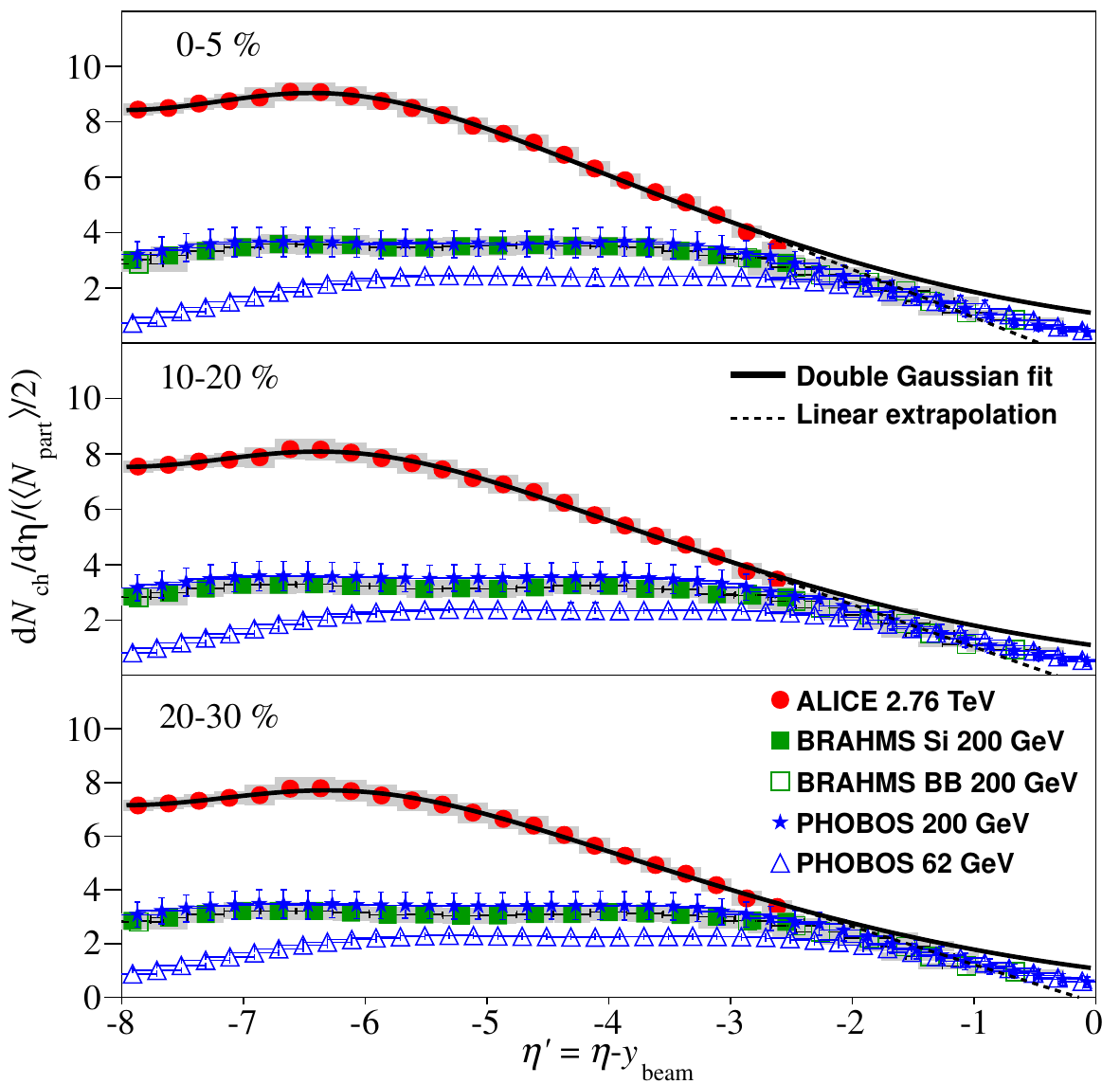}  
  \caption{The charged-particle pseudorapidity density distribution per participating nucleon pair
  for three centrality bins shown in the rest frame of one of the projectiles by using the variable
  $\eta' = \eta - y_{\mathrm{beam}}$ ($y_{\mathrm{beam}} = 7.99$). The \ALICE{} results are extrapolated to further
  values of $\eta'$ by fitting to the difference of two Gaussian functions (described earlier) and
  fitting a straight line to the last few points on the curve. These results are compared to lower
  energy data \protect\cite{Alver:2010ck,Bearden:2001qq}.}
  \label{fig:dndeta_scaling}
\end{figure}

It is well established that up to \RHIC{} energies the particle production in the fragmentation region
is invariant with the beam energy \cite{Benecke:1969sh}. This phenomenon is usually referred to as longitudinal scaling
and is observed by plotting the particle yields with respect to the variable $\eta - y_{\mathrm{beam}}$ \cite{Back:2002wb}. As it can be seen
from \figabbrevname~\ref{fig:dndeta_scaling}, our measurement is consistent with the validity of longitudinal scaling within the
errors arising mainly from the extrapolation of the charged-particle pseudorapidity density from the measured
region to the rapidity region of the projectile.

\begin{figure}[htbp]
  \centering
  \includegraphics[keepaspectratio,width=\figwidth]{%
    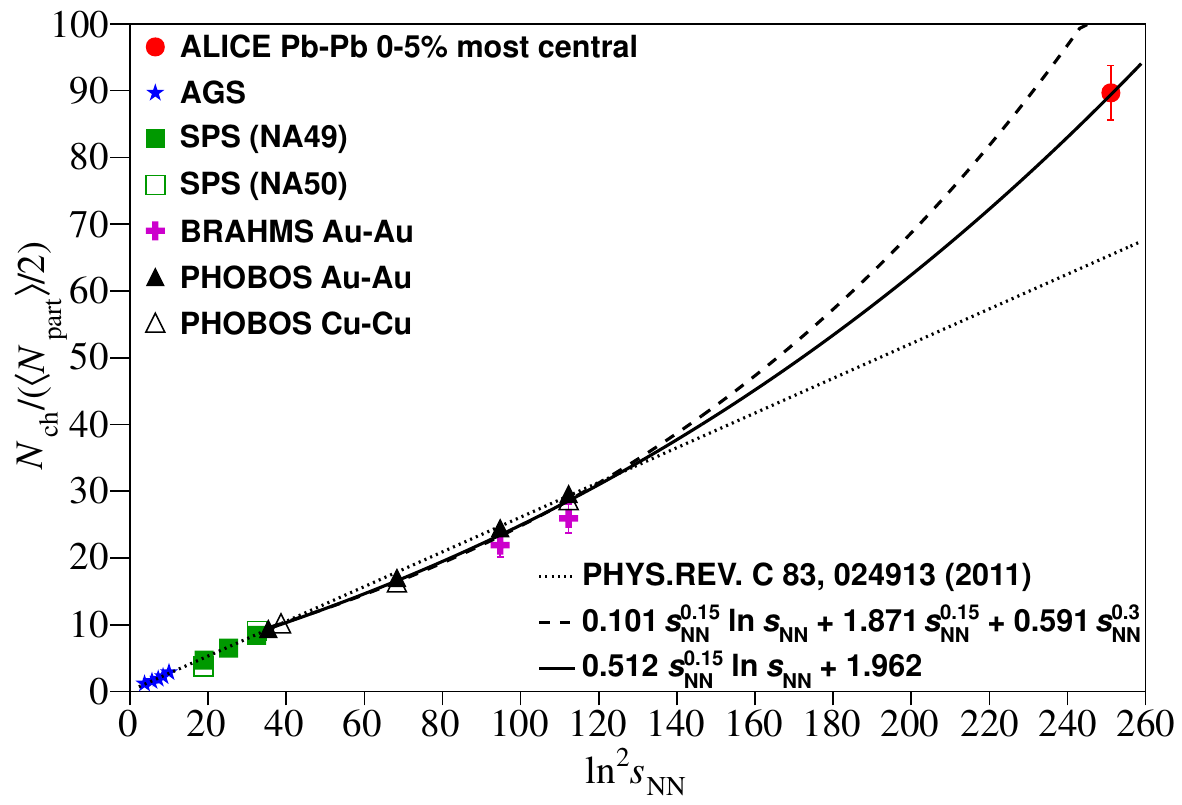}
  \caption{Total \Nch{} per participant pair versus $\ln^{2} \SNN{}$. Also shown (dotted
  line) is the fit to the \AGS{} \protect\cite{Klay:2003zf,Ahle:1998jc}, \SPS{} \protect\cite{Afanasiev:2002mx,Abreu:2002fw}, and \RHIC{}
  \protect\cite{Alver:2010ck,Bearden:2001xw,Bearden:2001qq} data from \protect\cite{Alver:2010ck} using the trapezoidal approximation
  for \dNdeta{} and assuming that the mid-rapidity \dNdeta{} scales as $\ln \SNN{}$. The dashed line is a fit to the
  \RHIC{} and \ALICE{} data derived using the trapezoidal approximation, but assuming the mid-rapidity
  \dNdeta{} scales as $\SNN[0.15]{}$ as in \protect\cite{Aamodt:2010pb}. The full drawn line is a fit to the
  \RHIC{} and \ALICE{} data derived assuming that \dNdeta{} is dominated by a flat mid-rapidity
  region with a width that grows as $\ln \SNN{}$.}
  \label{fig:nchln2s}
\end{figure}

The number of produced charged particles per participant pair was observed to have a linear
dependence on $\ln^{2} \SNN{}$ from \AGS{} to \RHIC{} energies based on a trapezoidal approximation for the
\dNdeta{} distribution with \dNdeta{} at mid-rapidity increasing proportional to $\ln \SNN{}$ \cite{Alver:2010ck}.
\figfullname~\ref{fig:nchln2s} shows this trend together with
the present \ALICE{} measurement. The trend does not persist to \LHC{} energies
and underpredicts the total number of produced charged particles at \sqrtSNNTeV{2.76}. To test if the trapezoidal approximation
for the \dNdeta{} distribution is still valid using a power law scaling of the mid-rapidity \dNdeta \cite{Aamodt:2010pb},
a new fit was performed to the \RHIC{} and \ALICE{} data, but was found to overpredict the total number of produced charged particles
at \sqrtSNNTeV{2.76}. Therefore, the trapezoidal approximation does not hold to \LHC{} energies. Instead, a fit with
a mid-rapidity \dNdeta{} value that scales as a power law as in \cite{Aamodt:2010pb} and extends over an $\eta$ range
scaling with $\ln \SNN{}$ gives a better general description.

\begin{figure}[htbp]
  \centering
  \includegraphics[keepaspectratio,width=\figwidth]{%
    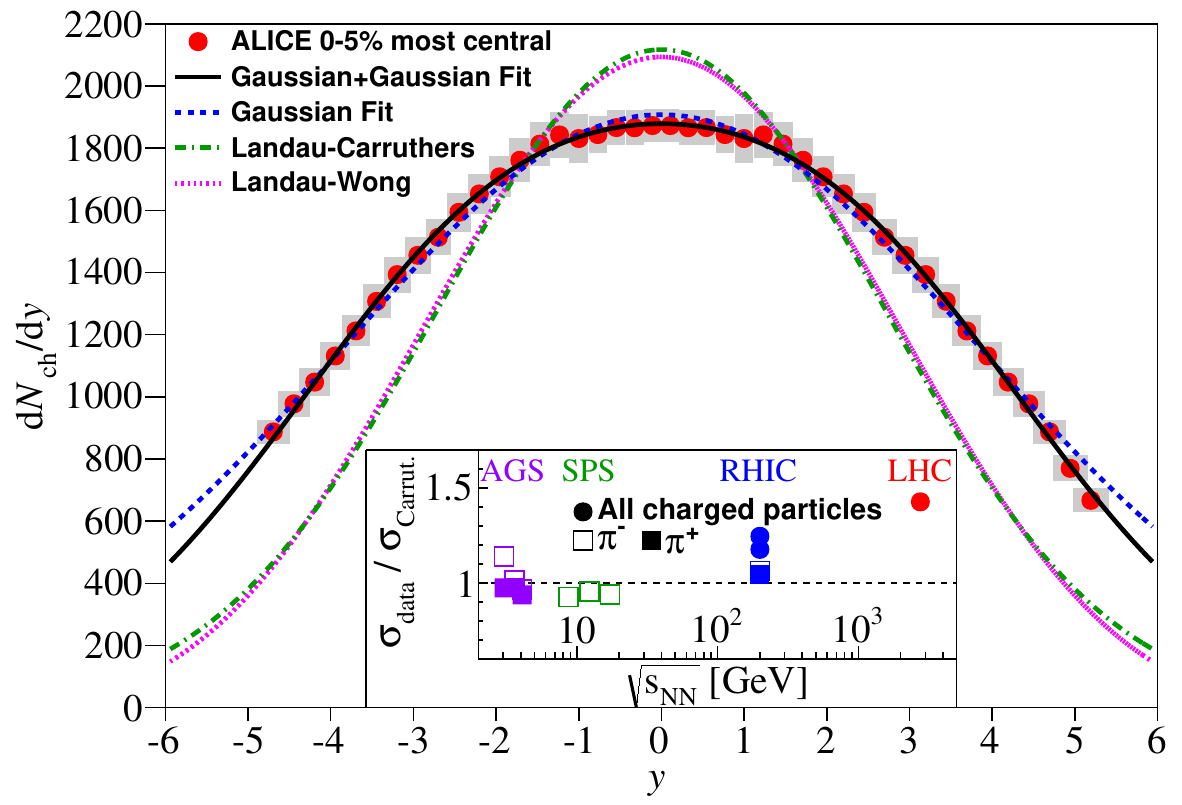}
  \caption{\dNdy{} distribution for the 5\% most central \PbPb{} collisions. A Gaussian distribution has been fit
  to the data ($\sigma = 3.86$). A Landau-Carruthers Gaussian \protect\cite{Carruthers:1973rw} and a Landau-Wong function
  \protect\cite{Wong:2008ex} are also shown.
  The full drawn line shows a fit to the sum of two Gaussian distributions of equal widths
  with the means at $\eta = \pm 2.17$ and $\sigma = 2.6$ as its area reproduces the estimated total number of charged particles.
  The inset shows the energy dependence of the ratio of $\sigma$ from a Gaussian fit
  to the expected Landau-Carruthers $\sigma$ taken from \protect\cite{Bearden:2004yx} extended to \sqrtSNNTeV{2.76}
  along with including \RHIC{} points derived from the \dNdeta{} distributions measured by \BRAHMS{} \cite{Bearden:2001qq}
  and \PHOBOS{} \cite{Back:2002wb} converted to \dNdy{} (the higher point and the lower point, respectively) using the same
  method employed at \sqrtSNNTeV{2.76}.}
  \label{fig:dndy}
\end{figure}

\figfullname~\ref{fig:dndy} shows the \dNdy{} distribution versus $y$ estimated by performing a Jacobian transformation from $\eta$ to
$y$ utilizing the measured particle ratios and \pt{} distributions in \ALICE{} for $\pi^{\pm}$, $K^{\pm}$, $p$,
and $\bar{p}$ at mid-rapidity \cite{Abelev:2012wca}.
The systematic error on the estimate includes a linear softening of the \pt{} spectra with $|\eta|$ where the $\langle \pt{} \rangle$
at $\eta = 3$ is $0.8$ of the $\langle \pt{} \rangle$ at $\eta = 0$ corresponding to approximately twice that seen for pions at \RHIC{}
\cite{Bearden:2004yx}. The systematic error also includes variations in the particle ratios of $\pm 50$\% beyond $\eta = 2.5$ and
a linear reduction in these variations to $0$ as $\eta \rightarrow 0$.
The contribution from net-protons was neglected as they contribute predominately near beam rapidity
and was, therefore, considered small relative to the variations in the other parameters.
While the data, within systematic errors, are consistent with a flat rapidity
plateau of about $\pm 1.5$ units around $y = 0$, they are also well described over the full
acceptance by a wide Gaussian distribution with $\sigma = 3.86$. This width, however, is
larger than expected from Landau hydrodynamics \cite{Carruthers:1973rw,Wong:2008ex}.
Lower-energy distributions derived
from identified pions have a width much closer to that expected from Landau hydrodynamics (see inset in
\figabbrevname~\ref{fig:dndy}). Two measurements derived from charged particles were computed using the
\pt{} spectra and particle ratios measured by \STAR{} \cite{Adams:2003xp} to convert the \dNdeta{} distributions
measured by \BRAHMS{} \cite{Bearden:2001qq} and \PHOBOS{} \cite{Back:2002wb} to \dNdy{} distributions in the same
way as previously applied to the \ALICE{} measurement. While the widths are larger than those derived from
identified pions at \sqrtSNNGeV{200}, there remains a significant increase from \RHIC{} to \LHC{} energies.
Similar observations of deviations from Landau hydrodynamics have been seen in other \PbPb{} measurements at
\sqrtSNNTeV{2.76} \cite{Chatrchyan:2012mb}.

\section{Conclusions}

The charged-particle pseudorapidity density distribution has been measured in \PbPb{}
collisions at \sqrtSNNTeV{2.76}. Results were obtained using a special sample of triggered
`satellite' collisions which allowed for reliable multiplicity measurements in the 0--30\%
centrality range. The measurement was performed in a wide pseudorapidity interval of
$-5.0 < \eta < 5.5$ allowing for the first estimate of the total number of charged particles
produced at the \LHC{}. The available theoretical predictions do not describe the data
satisfactorily although the CGC based model does well within its limited pseudorapidity
range. We do not observe a significant change in the shape of the distributions as a
function of the event centrality. Our results are compatible with the
preservation of longitudinal scaling up to \sqrtSNNTeV{2.76}. The scaling of total
number of charged particles per participant pair with $\ln^{2} \SNN{}$ does not persist to
\LHC{} energies. The \dNdy{} distribution of particles has a much larger width than that expected from
Landau hydrodynamics, showing an increasing deviation at higher energies.
%
\newenvironment{acknowledgement}{\relax}{\relax}
\begin{acknowledgement}
\section*{Acknowledgements}
The ALICE collaboration would like to thank all its engineers and technicians for their invaluable contributions to the construction of the experiment and the CERN accelerator teams for the outstanding performance of the LHC complex.
The ALICE collaboration acknowledges the following funding agencies for their support in building and
running the ALICE detector:
State Committee of Science,  World Federation of Scientists (WFS)
and Swiss Fonds Kidagan, Armenia,
Conselho Nacional de Desenvolvimento Cient\'{\i}fico e Tecnol\'{o}gico (CNPq), Financiadora de Estudos e Projetos (FINEP),
Funda\c{c}\~{a}o de Amparo \`{a} Pesquisa do Estado de S\~{a}o Paulo (FAPESP);
National Natural Science Foundation of China (NSFC), the Chinese Ministry of Education (CMOE)
and the Ministry of Science and Technology of China (MSTC);
Ministry of Education and Youth of the Czech Republic;
Danish Natural Science Research Council, the Carlsberg Foundation and the Danish National Research Foundation;
The European Research Council under the European Community's Seventh Framework Programme;
Helsinki Institute of Physics and the Academy of Finland;
French CNRS-IN2P3, the `Region Pays de Loire', `Region Alsace', `Region Auvergne' and CEA, France;
German BMBF and the Helmholtz Association;
General Secretariat for Research and Technology, Ministry of
Development, Greece;
Hungarian OTKA and National Office for Research and Technology (NKTH);
Department of Atomic Energy and Department of Science and Technology of the Government of India;
Istituto Nazionale di Fisica Nucleare (INFN) and Centro Fermi -
Museo Storico della Fisica e Centro Studi e Ricerche "Enrico
Fermi", Italy;
MEXT Grant-in-Aid for Specially Promoted Research, Ja\-pan;
Joint Institute for Nuclear Research, Dubna;
National Research Foundation of Korea (NRF);
CONACYT, DGAPA, M\'{e}xico, ALFA-EC and the EPLANET Program
(European Particle Physics Latin American Network)
Stichting voor Fundamenteel Onderzoek der Materie (FOM) and the Nederlandse Organisatie voor Wetenschappelijk Onderzoek (NWO), Netherlands;
Research Council of Norway (NFR);
Polish Ministry of Science and Higher Education;
National Authority for Scientific Research - NASR (Autoritatea Na\c{t}ional\u{a} pentru Cercetare \c{S}tiin\c{t}ific\u{a} - ANCS);
Ministry of Education and Science of Russian Federation, Russian
Academy of Sciences, Russian Federal Agency of Atomic Energy,
Russian Federal Agency for Science and Innovations and The Russian
Foundation for Basic Research;
Ministry of Education of Slovakia;
Department of Science and Technology, South Africa;
CIEMAT, EELA, Ministerio de Econom\'{i}a y Competitividad (MINECO) of Spain, Xunta de Galicia (Conseller\'{\i}a de Educaci\'{o}n),
CEA\-DEN, Cubaenerg\'{\i}a, Cuba, and IAEA (International Atomic Energy Agency);
Swedish Research Council (VR) and Knut $\&$ Alice Wallenberg
Foundation (KAW);
Ukraine Ministry of Education and Science;
United Kingdom Science and Technology Facilities Council (STFC);
The United States Department of Energy, the United States National
Science Foundation, the State of Texas, and the State of Ohio.
\end{acknowledgement}
\bibliographystyle{apsrev4-1}
\bibliography{Centrality_dependence_of_pseudorapidity_distribution}              
\newpage
%
%
\appendix
\section{The ALICE Collaboration}
\label{app:collab}

\begingroup
\small
\begin{flushleft}
E.~Abbas\Irefn{org36632}\And
B.~Abelev\Irefn{org1234}\And
J.~Adam\Irefn{org1274}\And
D.~Adamov\'{a}\Irefn{org1283}\And
A.M.~Adare\Irefn{org1260}\And
M.M.~Aggarwal\Irefn{org1157}\And
G.~Aglieri~Rinella\Irefn{org1192}\And
M.~Agnello\Irefn{org1313}\textsuperscript{,}\Irefn{org1017688}\And
A.G.~Agocs\Irefn{org1143}\And
A.~Agostinelli\Irefn{org1132}
% ...
\renewcommand\labelenumi{\textsuperscript{\theenumi}~}
\section*{Affiliation notes}
\renewcommand\theenumi{\roman{enumi}}
\begin{Authlist}
\item \Adef{M.V.Lomonosov Moscow State University, D.V.Skobeltsyn Institute of Nuclear Physics, Moscow, Russia}Also at: M.V.Lomonosov Moscow State University, D.V.Skobeltsyn Institute of Nuclear Physics, Moscow, Russia
\item \Adef{University of Belgrade, Faculty of Physics and Vinca Institute of Nuclear Sciences, Belgrade, Serbia}Also at: University of Belgrade, Faculty of Physics and Vinca Institute of Nuclear Sciences, Belgrade, Serbia
\item \Adef{Institute of Theoretical Physics, University of Wroclaw, Wroclaw, Poland}Also at: Institute of Theoretical Physics, University of Wroclaw, Wroclaw, Poland
\end{Authlist}
\section*{Collaboration Institutes}
\renewcommand\theenumi{\arabic{enumi}~}
\begin{Authlist}
\item \Idef{org36632}Academy of Scientific Research and Technology (ASRT), Cairo, Egypt
\item \Idef{org1332}A. I. Alikhanyan National Science Laboratory (Yerevan Physics Institute) Foundation, Yerevan, Armenia
\item \Idef{org1279}Benem\'{e}rita Universidad Aut\'{o}noma de Puebla, Puebla, Mexico
\item \Idef{org1220}Bogolyubov Institute for Theoretical Physics, Kiev, Ukraine
\item \Idef{org20959}Bose Institute, Department of Physics and Centre for Astroparticle Physics and Space Science (CAPSS), Kolkata, India
\item \Idef{org1262}Budker Institute for Nuclear Physics, Novosibirsk, Russia
\item \Idef{org1292}California Polytechnic State University, San Luis Obispo, California, United States
\item \Idef{org1329}Central China Normal University, Wuhan, China
\item \Idef{org14939}Centre de Calcul de l'IN2P3, Villeurbanne, France
\item \Idef{org1197}Centro de Aplicaciones Tecnol\'{o}gicas y Desarrollo Nuclear (CEADEN), Havana, Cuba
\item \Idef{org1242}Centro de Investigaciones Energ\'{e}ticas Medioambientales y Tecnol\'{o}gicas (CIEMAT), Madrid, Spain
\item \Idef{org1244}Centro de Investigaci\'{o}n y de Estudios Avanzados (CINVESTAV), Mexico City and M\'{e}rida, Mexico
\item \Idef{org1335}Centro Fermi - Museo Storico della Fisica e Centro Studi e Ricerche ``Enrico Fermi'', Rome, Italy
\item \Idef{org17347}Chicago State University, Chicago, United States
\item \Idef{org1288}Commissariat \`{a} l'Energie Atomique, IRFU, Saclay, France
\item \Idef{org15782}COMSATS Institute of Information Technology (CIIT), Islamabad, Pakistan
\item \Idef{org1294}Departamento de F\'{\i}sica de Part\'{\i}culas and IGFAE, Universidad de Santiago de Compostela, Santiago de Compostela, Spain
\item \Idef{org1106}Department of Physics Aligarh Muslim University, Aligarh, India
\item \Idef{org1121}Department of Physics and Technology, University of Bergen, Bergen, Norway
\item \Idef{org1162}Department of Physics, Ohio State University, Columbus, Ohio, United States
\item \Idef{org1300}Department of Physics, Sejong University, Seoul, South Korea
\item \Idef{org1268}Department of Physics, University of Oslo, Oslo, Norway
\item \Idef{org1315}Dipartimento di Fisica dell'Universit\`{a} and Sezione INFN, Trieste, Italy
\item \Idef{org1145}Dipartimento di Fisica dell'Universit\`{a} and Sezione INFN, Cagliari, Italy
\item \Idef{org1312}Dipartimento di Fisica dell'Universit\`{a} and Sezione INFN, Turin, Italy
\item \Idef{org1285}Dipartimento di Fisica dell'Universit\`{a} `La Sapienza' and Sezione INFN, Rome, Italy
\item \Idef{org1154}Dipartimento di Fisica e Astronomia dell'Universit\`{a} and Sezione INFN, Catania, Italy
\item \Idef{org1132}Dipartimento di Fisica e Astronomia dell'Universit\`{a} and Sezione INFN, Bologna, Italy
\item \Idef{org1270}Dipartimento di Fisica e Astronomia dell'Universit\`{a} and Sezione INFN, Padova, Italy
\item \Idef{org1290}Dipartimento di Fisica `E.R.~Caianiello' dell'Universit\`{a} and Gruppo Collegato INFN, Salerno, Italy
\item \Idef{org1103}Dipartimento di Scienze e Innovazione Tecnologica dell'Universit\`{a} del Piemonte Orientale and Gruppo Collegato INFN, Alessandria, Italy
\item \Idef{org1114}Dipartimento Interateneo di Fisica `M.~Merlin' and Sezione INFN, Bari, Italy
\item \Idef{org1237}Division of Experimental High Energy Physics, University of Lund, Lund, Sweden
\item \Idef{org1192}European Organization for Nuclear Research (CERN), Geneva, Switzerland
\item \Idef{org1227}Fachhochschule K\"{o}ln, K\"{o}ln, Germany
\item \Idef{org1122}Faculty of Engineering, Bergen University College, Bergen, Norway
\item \Idef{org1136}Faculty of Mathematics, Physics and Informatics, Comenius University, Bratislava, Slovakia
\item \Idef{org1274}Faculty of Nuclear Sciences and Physical Engineering, Czech Technical University in Prague, Prague, Czech Republic
\item \Idef{org1229}Faculty of Science, P.J.~\v{S}af\'{a}rik University, Ko\v{s}ice, Slovakia
\item \Idef{org1184}Frankfurt Institute for Advanced Studies, Johann Wolfgang Goethe-Universit\"{a}t Frankfurt, Frankfurt, Germany
\item \Idef{org1215}Gangneung-Wonju National University, Gangneung, South Korea
\item \Idef{org20958}Gauhati University, Department of Physics, Guwahati, India
\item \Idef{org1212}Helsinki Institute of Physics (HIP) and University of Jyv\"{a}skyl\"{a}, Jyv\"{a}skyl\"{a}, Finland
\item \Idef{org1203}Hiroshima University, Hiroshima, Japan
\item \Idef{org1254}Indian Institute of Technology Bombay (IIT), Mumbai, India
\item \Idef{org36378}Indian Institute of Technology Indore, Indore, India (IITI)
\item \Idef{org1266}Institut de Physique Nucl\'{e}aire d'Orsay (IPNO), Universit\'{e} Paris-Sud, CNRS-IN2P3, Orsay, France
\item \Idef{org1277}Institute for High Energy Physics, Protvino, Russia
\item \Idef{org1249}Institute for Nuclear Research, Academy of Sciences, Moscow, Russia
\item \Idef{org1320}Nikhef, National Institute for Subatomic Physics and Institute for Subatomic Physics of Utrecht University, Utrecht, Netherlands
\item \Idef{org1250}Institute for Theoretical and Experimental Physics, Moscow, Russia
\item \Idef{org1230}Institute of Experimental Physics, Slovak Academy of Sciences, Ko\v{s}ice, Slovakia
\item \Idef{org1127}Institute of Physics, Bhubaneswar, India
\item \Idef{org1275}Institute of Physics, Academy of Sciences of the Czech Republic, Prague, Czech Republic
\item \Idef{org1139}Institute of Space Sciences (ISS), Bucharest, Romania
\item \Idef{org27399}Institut f\"{u}r Informatik, Johann Wolfgang Goethe-Universit\"{a}t Frankfurt, Frankfurt, Germany
\item \Idef{org1185}Institut f\"{u}r Kernphysik, Johann Wolfgang Goethe-Universit\"{a}t Frankfurt, Frankfurt, Germany
\item \Idef{org1177}Institut f\"{u}r Kernphysik, Technische Universit\"{a}t Darmstadt, Darmstadt, Germany
\item \Idef{org1256}Institut f\"{u}r Kernphysik, Westf\"{a}lische Wilhelms-Universit\"{a}t M\"{u}nster, M\"{u}nster, Germany
\item \Idef{org1246}Instituto de Ciencias Nucleares, Universidad Nacional Aut\'{o}noma de M\'{e}xico, Mexico City, Mexico
\item \Idef{org1247}Instituto de F\'{\i}sica, Universidad Nacional Aut\'{o}noma de M\'{e}xico, Mexico City, Mexico
\item \Idef{org1308}Institut Pluridisciplinaire Hubert Curien (IPHC), Universit\'{e} de Strasbourg, CNRS-IN2P3, Strasbourg, France
\item \Idef{org1182}Joint Institute for Nuclear Research (JINR), Dubna, Russia
\item \Idef{org1199}Kirchhoff-Institut f\"{u}r Physik, Ruprecht-Karls-Universit\"{a}t Heidelberg, Heidelberg, Germany
\item \Idef{org20954}Korea Institute of Science and Technology Information, Daejeon, South Korea
\item \Idef{org1017642}KTO Karatay University, Konya, Turkey
\item \Idef{org1160}Laboratoire de Physique Corpusculaire (LPC), Clermont Universit\'{e}, Universit\'{e} Blaise Pascal, CNRS--IN2P3, Clermont-Ferrand, France
\item \Idef{org1194}Laboratoire de Physique Subatomique et de Cosmologie (LPSC), Universit\'{e} Joseph Fourier, CNRS-IN2P3, Institut Polytechnique de Grenoble, Grenoble, France
\item \Idef{org1187}Laboratori Nazionali di Frascati, INFN, Frascati, Italy
\item \Idef{org1232}Laboratori Nazionali di Legnaro, INFN, Legnaro, Italy
\item \Idef{org1125}Lawrence Berkeley National Laboratory, Berkeley, California, United States
\item \Idef{org1234}Lawrence Livermore National Laboratory, Livermore, California, United States
\item \Idef{org1251}Moscow Engineering Physics Institute, Moscow, Russia
\item \Idef{org1322}National Centre for Nuclear Studies, Warsaw, Poland
\item \Idef{org1140}National Institute for Physics and Nuclear Engineering, Bucharest, Romania
\item \Idef{org1017626}National Institute of Science Education and Research, Bhubaneswar, India
\item \Idef{org1165}Niels Bohr Institute, University of Copenhagen, Copenhagen, Denmark
\item \Idef{org1109}Nikhef, National Institute for Subatomic Physics, Amsterdam, Netherlands
\item \Idef{org1283}Nuclear Physics Institute, Academy of Sciences of the Czech Republic, \v{R}e\v{z} u Prahy, Czech Republic
\item \Idef{org1264}Oak Ridge National Laboratory, Oak Ridge, Tennessee, United States
\item \Idef{org1189}Petersburg Nuclear Physics Institute, Gatchina, Russia
\item \Idef{org1170}Physics Department, Creighton University, Omaha, Nebraska, United States
\item \Idef{org1157}Physics Department, Panjab University, Chandigarh, India
\item \Idef{org1112}Physics Department, University of Athens, Athens, Greece
\item \Idef{org1152}Physics Department, University of Cape Town and  iThemba LABS, National Research Foundation, Somerset West, South Africa
\item \Idef{org1209}Physics Department, University of Jammu, Jammu, India
\item \Idef{org1207}Physics Department, University of Rajasthan, Jaipur, India
\item \Idef{org1200}Physikalisches Institut, Ruprecht-Karls-Universit\"{a}t Heidelberg, Heidelberg, Germany
\item \Idef{org1017688}Politecnico di Torino, Turin, Italy
\item \Idef{org1325}Purdue University, West Lafayette, Indiana, United States
\item \Idef{org1281}Pusan National University, Pusan, South Korea
\item \Idef{org1176}Research Division and ExtreMe Matter Institute EMMI, GSI Helmholtzzentrum f\"ur Schwerionenforschung, Darmstadt, Germany
\item \Idef{org1334}Rudjer Bo\v{s}kovi\'{c} Institute, Zagreb, Croatia
\item \Idef{org1298}Russian Federal Nuclear Center (VNIIEF), Sarov, Russia
\item \Idef{org1252}Russian Research Centre Kurchatov Institute, Moscow, Russia
\item \Idef{org1224}Saha Institute of Nuclear Physics, Kolkata, India
\item \Idef{org1130}School of Physics and Astronomy, University of Birmingham, Birmingham, United Kingdom
\item \Idef{org1338}Secci\'{o}n F\'{\i}sica, Departamento de Ciencias, Pontificia Universidad Cat\'{o}lica del Per\'{u}, Lima, Peru
\item \Idef{org1155}Sezione INFN, Catania, Italy
\item \Idef{org1313}Sezione INFN, Turin, Italy
\item \Idef{org1271}Sezione INFN, Padova, Italy
\item \Idef{org1133}Sezione INFN, Bologna, Italy
\item \Idef{org1146}Sezione INFN, Cagliari, Italy
\item \Idef{org1316}Sezione INFN, Trieste, Italy
\item \Idef{org1115}Sezione INFN, Bari, Italy
\item \Idef{org1286}Sezione INFN, Rome, Italy
\item \Idef{org36377}Nuclear Physics Group, STFC Daresbury Laboratory, Daresbury, United Kingdom
\item \Idef{org1258}SUBATECH, Ecole des Mines de Nantes, Universit\'{e} de Nantes, CNRS-IN2P3, Nantes, France
\item \Idef{org35706}Suranaree University of Technology, Nakhon Ratchasima, Thailand
\item \Idef{org1304}Technical University of Split FESB, Split, Croatia
\item \Idef{org1017659}Technische Universit\"{a}t M\"{u}nchen, Munich, Germany
\item \Idef{org1168}The Henryk Niewodniczanski Institute of Nuclear Physics, Polish Academy of Sciences, Cracow, Poland
\item \Idef{org17361}The University of Texas at Austin, Physics Department, Austin, TX, United States
\item \Idef{org1173}Universidad Aut\'{o}noma de Sinaloa, Culiac\'{a}n, Mexico
\item \Idef{org1296}Universidade de S\~{a}o Paulo (USP), S\~{a}o Paulo, Brazil
\item \Idef{org1149}Universidade Estadual de Campinas (UNICAMP), Campinas, Brazil
\item \Idef{org1239}Universit\'{e} de Lyon, Universit\'{e} Lyon 1, CNRS/IN2P3, IPN-Lyon, Villeurbanne, France
\item \Idef{org1205}University of Houston, Houston, Texas, United States
\item \Idef{org20371}University of Technology and Austrian Academy of Sciences, Vienna, Austria
\item \Idef{org1222}University of Tennessee, Knoxville, Tennessee, United States
\item \Idef{org1310}University of Tokyo, Tokyo, Japan
\item \Idef{org1318}University of Tsukuba, Tsukuba, Japan
\item \Idef{org21360}Eberhard Karls Universit\"{a}t T\"{u}bingen, T\"{u}bingen, Germany
\item \Idef{org1225}Variable Energy Cyclotron Centre, Kolkata, India
\item \Idef{org1017687}Vestfold University College, Tonsberg, Norway
\item \Idef{org1306}V.~Fock Institute for Physics, St. Petersburg State University, St. Petersburg, Russia
\item \Idef{org1323}Warsaw University of Technology, Warsaw, Poland
\item \Idef{org1179}Wayne State University, Detroit, Michigan, United States
\item \Idef{org1143}Wigner Research Centre for Physics, Hungarian Academy of Sciences, Budapest, Hungary
\item \Idef{org1260}Yale University, New Haven, Connecticut, United States
\item \Idef{org15649}Yildiz Technical University, Istanbul, Turkey
\item \Idef{org1301}Yonsei University, Seoul, South Korea
\item \Idef{org1327}Zentrum f\"{u}r Technologietransfer und Telekommunikation (ZTT), Fachhochschule Worms, Worms, Germany
\end{Authlist}
\endgroup

%
\end{document}